\documentstyle[12pt]{article}
\input psfig.sty
\def\nn{\noindent}

\def\Re{{\cal R \mskip-4mu \lower.1ex \hbox{\it e}\,}}
\def\Im{{\cal I \mskip-5mu \lower.1ex \hbox{\it m}\,}}
\def\ie{{\it i.e.}}
\def\eg{{\it e.g.}}

\def\etal{{\it et al.}}

\def\sub#1{_{\lower.25ex\hbox{$\scriptstyle#1$}}}

\def\to{\rightarrow}

\def\subw{_{\rm w}}
\def\mh{\ifmmode m\sbl H \else $m\sbl H$\fi}
\def\mch{\ifmmode m_{H^\pm} \else $m_{H^\pm}$\fi}
\def\mt{\ifmmode m_t\else $m_t$\fi}
\def\mc{\ifmmode m_c\else $m_c$\fi}
\def\mz{\ifmmode M_Z\else $M_Z$\fi}
\def\mw{\ifmmode M_W\else $M_W$\fi}
\def\mws{\ifmmode M_W^2 \else $M_W^2$\fi}
\def\mhs{\ifmmode m_H^2 \else $m_H^2$\fi}   
\def\mzs{\ifmmode M_Z^2 \else $M_Z^2$\fi}
\def\mts{\ifmmode m_t^2 \else $m_t^2$\fi}
\def\mcs{\ifmmode m_c^2 \else $m_c^2$\fi}
\def\mchs{\ifmmode m_{H^\pm}^2 \else $m_{H^\pm}^2$\fi}
\def\ztwo{\ifmmode Z_2\else $Z_2$\fi}
\def\zone{\ifmmode Z_1\else $Z_1$\fi}
\def\mtwo{\ifmmode M_2\else $M_2$\fi}
\def\mone{\ifmmode M_1\else $M_1$\fi}
\def\tb{\ifmmode \tan\beta \else $\tan\beta$\fi}
\def\xw{\ifmmode x\subw\else $x\subw$\fi}
\def\ch{\ifmmode H^\pm \else $H^\pm$\fi}
\def\lum{\ifmmode {\cal L}\else ${\cal L}$\fi}
\def\inpb{\ifmmode {\rm pb}^{-1}\else ${\rm pb}^{-1}$\fi}
\def\infb{\ifmmode {\rm fb}^{-1}\else ${\rm fb}^{-1}$\fi}
\def\epem{\ifmmode e^+e^-\else $e^+e^-$\fi}
\def\ppb{\ifmmode \bar pp\else $\bar pp$\fi}
\def\bsg{\ifmmode B\to X_s\gamma\else $B\to X_s\gamma$\fi}
\def\bsll{\ifmmode B\to X_s\ell^+\ell^-\else $B\to X_s\ell^+\ell^-$\fi}
\def\bstt{\ifmmode B\to X_s\tau^+\tau^-\else $B\to X_s\tau^+\tau^-$\fi}

\def\sq{\ifmmode \tilde q\else $\tilde q$\fi}
\def\glu{\ifmmode \tilde g\else $\tilde g$\fi}
\def\mglu{\ifmmode m_{\tilde g}\else $m_{\tilde g}$\fi}
\def\msq{\ifmmode m_{\tilde q}\else $m_{\tilde q}$\fi}
\newskip\zatskip \zatskip=0pt plus0pt minus0pt
\def\matth{\mathsurround=0pt}
\def\lsim{\mathrel{\mathpalette\atversim<}}
\def\gsim{\mathrel{\mathpalette\atversim>}}
\def\atversim#1#2{\lower0.7ex\vbox{\baselineskip\zatskip\lineskip\zatskip
  \lineskiplimit 0pt\ialign{$\matth#1\hfil##\hfil$\crcr#2\crcr\sim\crcr}}}

\renewcommand{\thefootnote}{\fnsymbol{footnote}}

\begin{document} \begin{titlepage} 
\rightline{\vbox{\halign{&#\hfil\cr
&SLAC-PUB-7372\cr
&December 1996\cr}}}
\vspace{1in} 
\begin{center}

{\Large\bf
Constraints on Light Gluinos from Tevatron Dijet Data}
\footnote{Work supported by the Department of 
Energy, Contract DE-AC03-76SF00515 and the Natural Sciences and Engineering
Council of Canada}
\medskip

\normalsize 
{\large JoAnne L. Hewett and Thomas G. Rizzo} \\
\vskip .3cm
Stanford Linear Accelerator Center, Stanford CA 94309, USA \\
\vskip .3cm
and\\
\vskip .3cm
{\large Michael A. Doncheski }\\
\vskip .3cm
Ottawa Carleton Institute for Physics, Carleton University, Ottawa, Canada\\
Department of Physics\footnote{present address}, 
Pennsylvania State University, Mont Alto, PA 17237, USA\\
\vskip .3cm

\end{center}

\begin{abstract} 

The effects of light, long-lived gluinos on $2\to 2$ processes at hadron 
colliders are examined.  Such particles can mediate single squark resonant 
production via $q\glu\to\sq\to q\glu$ which
would significantly modify the dijet data sample.  We find that
squark masses in the range $130<m_{\tilde q}< 694, 595, 573$ GeV are
excluded for gluino masses of $0.4, 1.3, 5.0$ GeV from existing UA2 and
Tevatron data on dijet bump searches and angular distributions.  Run II
of the Tevatron has the capability of excluding this scenario for squark masses
up to $\sim 1$ TeV.

\end{abstract} 


\renewcommand{\thefootnote}{\arabic{footnote}} \end{titlepage} 


Supersymmetry is a compelling candidate for physics beyond the Standard
Model (SM) and has engrossed both the theoretical and experimental communities.
Most of the attention has been focused on the minimal version of supersymmetry 
(MSSM), however, many other incarnations of supersymmetry could exist. 
In most cases these non-minimal models can significantly alter 
supersymmetric phenomenology and the associated search strategies, and hence 
all consequences of such models must be examined before regions of 
supersymmetric parameter space can be positively excluded.
Here, we examine one such non-minimal case: the light gluino 
scenario.  In some models it is natural\cite{radmass} for gluinos to be much 
lighter than, \eg, squarks if they acquire their masses radiatively.  While 
several experiments presently cast doubt on the existence of the low mass gluino
window ($\mglu\lsim 5$ GeV), it has yet to be conclusively ruled out (or 
verified).  In fact, the experimental bounds on this possibility are 
surprisingly spotty and controversial as evidenced by the continual debate in 
the literature\cite{pdg}.  It is thus imperative to examine all implications of 
this hypothesis in order to quell this dispute.  In this work, we investigate 
an additional data sample which provides strong constraints on the light gluino
scenario, namely $2\to 2$ processes at high energy hadron colliders.  

The window for a very light gluino was pointed out\cite{gfar} many years 
ago and its effects have since been analyzed in a variety of processes.
A resurgence of interest in this scenario surfaced with the relatively recent
observation\cite{clavalphas} that an apparent discrepancy between the value 
of $\alpha_s$ measured from jet production at SLD and LEP and that discerned 
from low energy data is resolved by the slower running of $\alpha_s$ in the 
presence of light gluinos.  However, recent compilations\cite{phil} of various
determinations of $\alpha_s$ no longer show evidence of such a discrepancy,
within the errors, but also
claim that the precision of each individual measurement is such that any
anomalous effect up to the $\sim 5\%$ level may not be perceived.  The
most noticeable consequence of this model is that the
standard signals for gluino and squark production are modified in the
presence of light gluinos.  The bounds on the gluino mass, $\mglu>144-224$ GeV 
from the Tevatron\cite{cdfd0} (with the range being due to the assumed
relative sizes of the squark and gluino masses), are 
invalidated in this case as they
depend on the fact that the \glu\ is short-lived and decays with the 
characteristic missing energy signature.  Thus to be light, gluinos must be
long-lived and appear to hadronize as jets.  Since they are unable to appear
as free particles, light gluinos will indeed form hadrons, with the bound states
having longer lifetimes, and fragment in such a way as to mimic jets in
a high energy detector\cite{morefar}.  If kinematically allowed, the gluino
hadrons will eventually decay into a final state
containing jets $+\chi^0_1$, where $\chi^0_1$ is the lightest neutralino.
The crucial ingredient for detection is then the ability of the final state
$\chi^0_1$ to pass the detector's missing energy cuts, which depends, amongst
other things, on how the \glu\ hadron fragments.  It has been
estimated\cite{uaone} that for $\mglu\gsim 5$ GeV the \glu\ would have
been detected at UA1.  However, as the gluino mass decreases, the missing
energy signal disappears altogether.
Standard squark searches are also nullified in this model as now the primary
decay is $\sq\to q\glu$, which again, escapes searches based on missing 
energy.  In this case, the squark mass bounds are reduced to $\msq>M_Z/2$,
with the mass constraint being extended to $50-60$ GeV from precision
electroweak measurements at SLC/LEP\cite{bhatt}.  We expect LEP II to
strengthen the squark mass bound to $\gsim 80-85$ GeV.

We now discuss the results from a variety of light gluino searches.
At present, the least controversial bound on light gluinos is from a search
by CUSB\cite{cusb} for radiative $\Upsilon$ decays into bound states of 
gluinos.  They exclude the mass range $\sim 1.5 - 3.5$ GeV (regardless of the
gluino lifetime), where
the lower limit is approximate due to questions\cite{gfartwo} concerning the 
validity of perturbative QCD in this regime.  ARGUS\cite{argus} looked for 
secondary vertices from $\chi_b\to g\glu\glu$ with subsequent decay of the 
gluino bound states and constrained a small region in the gluino 
mass - lifetime parameter space; these results, however, also suffer\cite{pdg} 
from perturbative QCD uncertainties as well as those from fragmentation effects.
Beam dump experiments\cite{bdump} have looked for secondary vertices from the
decay of \glu\ hadrons and appear to disfavor light gluinos for restricted 
regions of the gluino lifetime, but these results depend on (i) assumptions
on the production cross sections of the gluino hadrons,  (ii) the value of 
the squark mass (iii) the interactions of the lightest color-singlet 
supersymmetric particle, $\chi^0_1$, with the detector, and (iv) \glu\ 
fragmentation effects and decay models.  Searches for new neutral particles
at Fermilab exclude\cite{neutral} $2<\mglu<4$ GeV for \glu\ lifetimes in excess
of $10^{-7}$ s.  Jet
angular distributions of decays of the $Z$ into four jets and precision
measurements of the QCD structure constants $C_{A,F}$ and $T_F$ have been
shown to be particularly sensitive to the existence of light 
gluinos\cite{hitoshi}, but critically depend\cite{gfarthr} on currently
uncalculated higher order QCD corrections and hence no firm conclusions can
presently be drawn. The $Z$ boson can decay into 2 gluinos,
however the branching fraction is small\cite{drees} ($B\sim 0.06\%$), and 
would be hidden underneath ordinary QCD events.  The detection of light
gluinos at HERA, through their effect on deep inelastic structure
functions\cite{pdfs} or via their production in the $3+1$ jet photoproduction 
cross section\cite{stirl}, have also been shown to be difficult.

In this study, we examine the effects of light, long-lived gluinos on dijet
production in hadronic collisions.  One would expect the influence of light 
\glu's to be large in such processes since they contribute at leading order
in perturbation theory.
It has been shown\cite{zvi}, however, that competing effects tend to suppress
their impact on the single jet inclusive $E_T$ spectrum.  Nonetheless,
we find that the influence of resonant squark production from the subprocess 
$q\glu\to\sq\to q\glu$ should not be neglected as it greatly modifies the 
dijet mass spectrum and places strong constraints on the light gluino window.
Our conclusions avoid some of the aforementioned difficulties in constraining
this scenario, as non-perturbative QCD effects are negligible at the energies 
considered here and our results are insensitive to a long \glu\ lifetime.
The essential ingredients of this model for our analysis are
(i) the evolution of $\alpha_s$ is modified by the inclusion of light
gluinos in the QCD $\beta$ function, (ii) long-lived gluinos in the final state
hadronize as jets, and (iii) light gluinos contribute a non-negligible partonic
content of the proton.  This introduces several new $2\to 2$ parton scattering
processes, as well as modifying the Altarelli-Parisi evolution of the 
parton densities.  Global fits of structure functions which include a light
gluino distribution have been performed\cite{pdfs}, and
it has been found that the NLO \glu\ 
parton distributions are roughly three (five) times larger than that of the 
strange quark at large (small) $x$ for very light gluinos, $\mglu\lsim 1.5$ GeV,
and carry $\sim 5\%$ of the proton's momentum fraction at large $Q^2$ for 
$\mglu=5$ GeV.

We now proceed with our calculation.  All $2\to 2$ subprocesses have been
evaluated; they naturally fall into three categories, (i) those of the SM,
$qq\to qq$\,, ~$q\bar q\to q\bar q, gg$\,, ~~$qg\to qg$, and $gg\to q\bar q,gg$,
(ii) all SM initiated $2\to 2$ processes with final state gluinos, 
$q\bar q, gg\to \glu\glu$,
and (iii) all gluino initiated processes, $q\glu\to q\glu$\,,
~$g\glu\to g\glu$, and $\glu\glu\to gg,\glu\glu$.  Note that resonant squark
production appears in the latter set.  Higher order $2\to 3$ processes,
including the new reactions\cite{clav} which produce $\sq+$jet and thus yield 
3 jet final states once the squark decays, have not been included.
The mass of the light gluino has also been neglected in the
evaluation of the subprocess cross sections as the results should not be
sensitive to \mglu\ at the energy scales considered here.  The parton 
distributions\cite{pdfs} of R\" uckl and Vogt have been used for $\mglu\lsim
1.5$ GeV and those of Roberts and Stirling for $\mglu=5$ GeV.  These values
of the gluino mass avoid all of the experimental constraints detailed above.
The change in the evolution of $\alpha_s$ has been taken into account by
fixing $\alpha_s(M_Z)$ to the world average value\cite{pdg} and then running 
it to the relevant scale using the appropriate 2-loop $\beta$ functions.  We 
note that the 3-loop light \glu\ $\beta$ functions have only recently been
determined\cite{levan}.

In evaluating the squark resonance contribution to the cross section, we 
have used the narrow width approximation, which is valid for $\Gamma/m\lsim 
0.1$ and hence is reliable in this case.  We have included a 10\% 
contribution to the squark width for potential non-dijet decays, \ie, 
$\Gamma_{\tilde q}=1.1\times \Gamma(\sq\to q\glu)$.  This is
conservative as dijet decays will be by far the dominant mode.  The 10\% figure
should cover the additional weak decays 
$\sq\to q \chi^0_{1-4}$ and $\sq\to q \chi^\pm_{1,2}$, whichever are
kinematically allowed, as they are expected to have 
small branching fractions of order $\lsim 1-2\%$ each and hence are suppressed
compared to the dijet mode.  We note that monojet signals from squark 
production in this scenario have been previously analyzed\cite{mono}.
We have also assumed that there are 5 degenerate squarks, with equal 
masses for the left- and right-handed states.  Our results are not dependent
on this assumption, however, as the contribution of each squark
flavor to the resonance peak is weighted by the corresponding quark's parton 
density.  Hence this supposition does not simply result in an overall 
multiplicative factor to the cross section.  In fact, the
charm and bottom squarks have essentially negligible contributions to the
resonance peak.

Experimentally, the dijet system consists of the two jets with the highest 
transverse momentum in the event.  In all cases, except where noted, we apply 
the cuts used by the CDF Collaboration\cite{cdfdj} in their dijet analyses.  
This corresponds to $p_{T_j}>20$ GeV, $|\eta_{1,2}|<2$, where $\eta_{1,2}$
are the pseudorapidities of the two leading jets, and $|\cos\theta^*|\leq 2/3$, 
with $\theta^*$ being the parton-parton scattering angle in the center
of mass frame.  Following CDF, we evaluate these processes at the scale 
$\mu=p_T$.  In Fig. 1 we display the dijet invariant mass and single
jet inclusive $p_T$ 
distributions for the cases of $\msq=300$ and 500 GeV, taking
$\mglu=0.4, 1.3$ and 5 GeV corresponding to the dotted, dashed, and solid 
curves, respectively.  In the case of the $p_T$ distributions, we assume 
$|\eta_{1,2}|<0.5$, $\mu=p_T/2$, and no angular cuts are applied.  We
see that the resonance peaks stand out for all values of the gluino mass.
Note the degradation of the cross section as the \glu\ mass increases.

We now evaluate the dijet resonance cross section and compare it
to searches for dijet mass peaks from the single 
production of new particles performed by hadron collider 
experiments\cite{cdfdj,ua2,d0dj}.  Figure 2 presents the single squark 
production cross section in the dijet channel as a function of the squark mass 
for various values of \mglu.  Also displayed in the figure (dotted curve) is 
the upper limit on the production of dijet resonances at (a) UA2\cite{ua2} at 
$90\%$ C.L., as well as both (b) CDF\cite{cdfdj} and (c) D0\cite{d0dj} 
at the $95\%$ C.L. In the D0 case the applied cuts are somewhat different than 
those employed by CDF: $|\eta_{1,2}|<1$ and $|\eta_1-\eta_2|<1.6$. 
We see that the three experiments combine
to exclude substantial  regions of the light gluino parameter 
space.  The ranges of the squark masses which are ruled out for each value of
\mglu\  are summarized in Table \ref{bumptab}.  We do not expect the bounds
to drastically improve as $\mglu\to 0$ as the 
squark resonance cross section is
not appreciably changing as the gluino mass decreases (once $\mglu\lsim 1.5$
GeV) as shown in Fig. 2. A short analysis shows that the cross section for 
massless gluinos is approximately 1.3(1.6) times larger than that for the case 
of $\mglu=0.4$ GeV at low(high) dijet invariant masses. 

\begin{table}
\centering
\begin{tabular}{|c|c|c|c|} \hline\hline
\mglu\ & \msq\ ~(UA2) & \msq\ ~(CDF) & \msq\ ~(D0) \\ \hline
$0.4-1.3$ & 130 to $195-220$ & 220 to 475 & 310 to 590\\
5.0 & 130 to 170 & 240 to 455 & 320 to 460\\ \hline\hline
\end{tabular}
\caption{The squark mass regions in GeV excluded by the searches for
dijet resonances by the UA2 (at $90\%$ C.L.), CDF and D0 (at $95\%$ C.L.) 
Collaborations for an assumed gluino mass.}
\label{bumptab}
\end{table}

\nn
\begin{figure}[htbp]
\centerline{
\psfig{figure=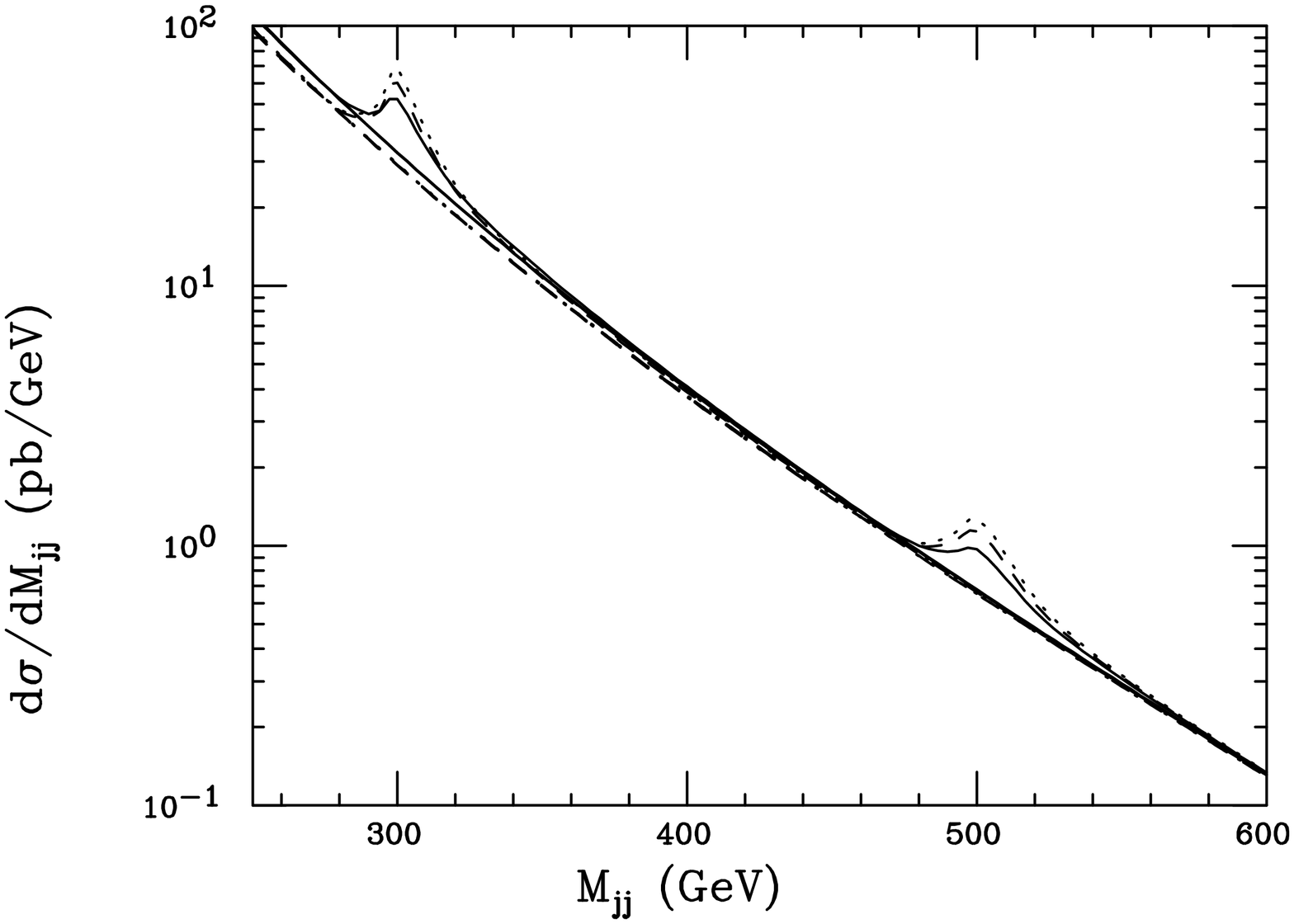,height=10cm,width=12cm,angle=-90}}
\vspace*{-0.75cm}
\centerline{
\psfig{figure=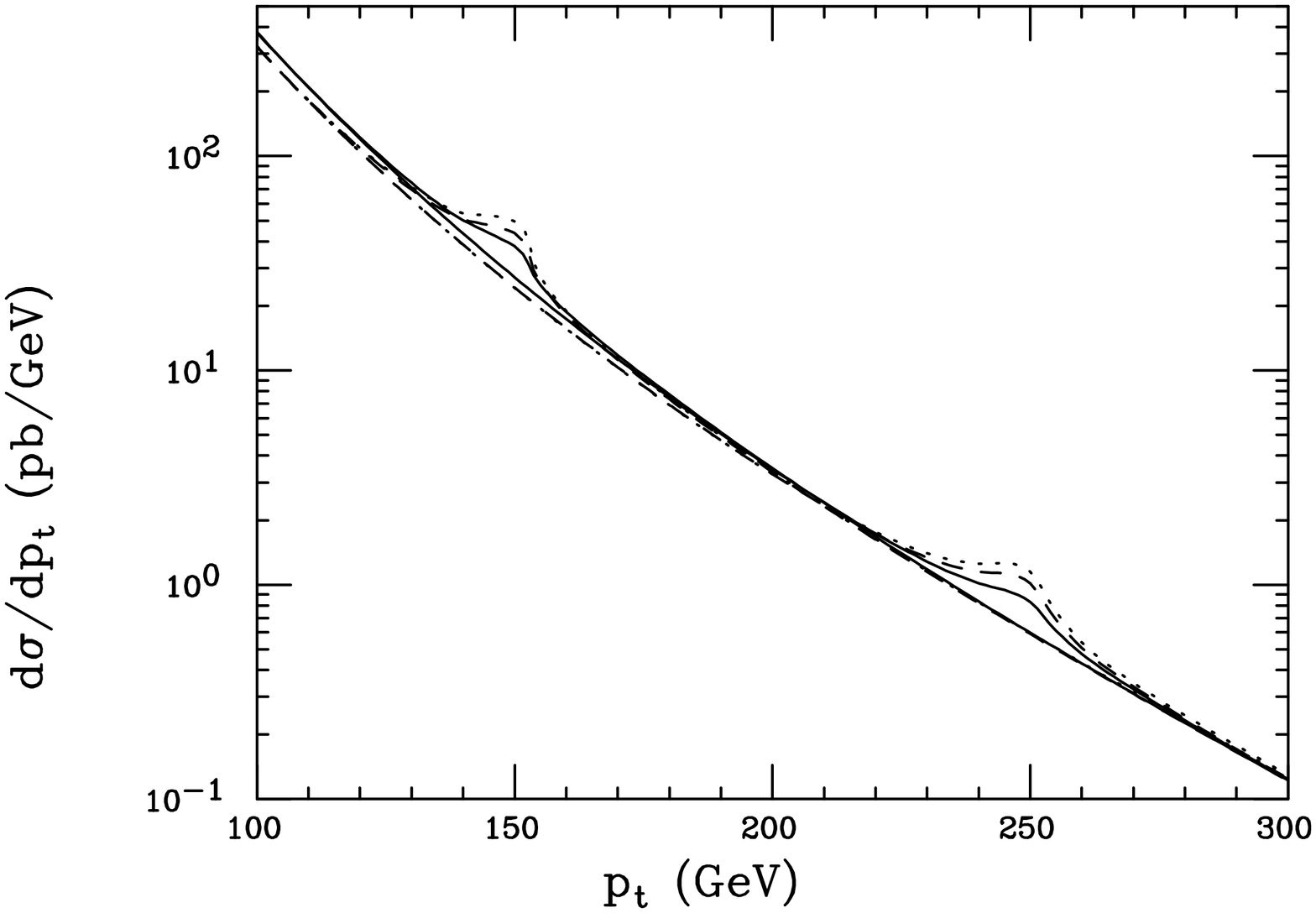,height=10cm,width=12cm,angle=-90}}
\vspace*{-0.75cm}
\caption{The (a) dijet invariant mass and (b) single jet inclusive $p_T$
distributions for the $2\to 2$ processes described in the text for the
two cases $\msq=300$, and 500 GeV.  The gluino mass is taken to be 0.4, 1.3,
and 5.0 GeV corresponding to the dotted, dashed, and solid curves, respectively.
}
\label{bumps}
\end{figure}

\nn
\begin{figure}[htbp]
\centerline{
\psfig{figure=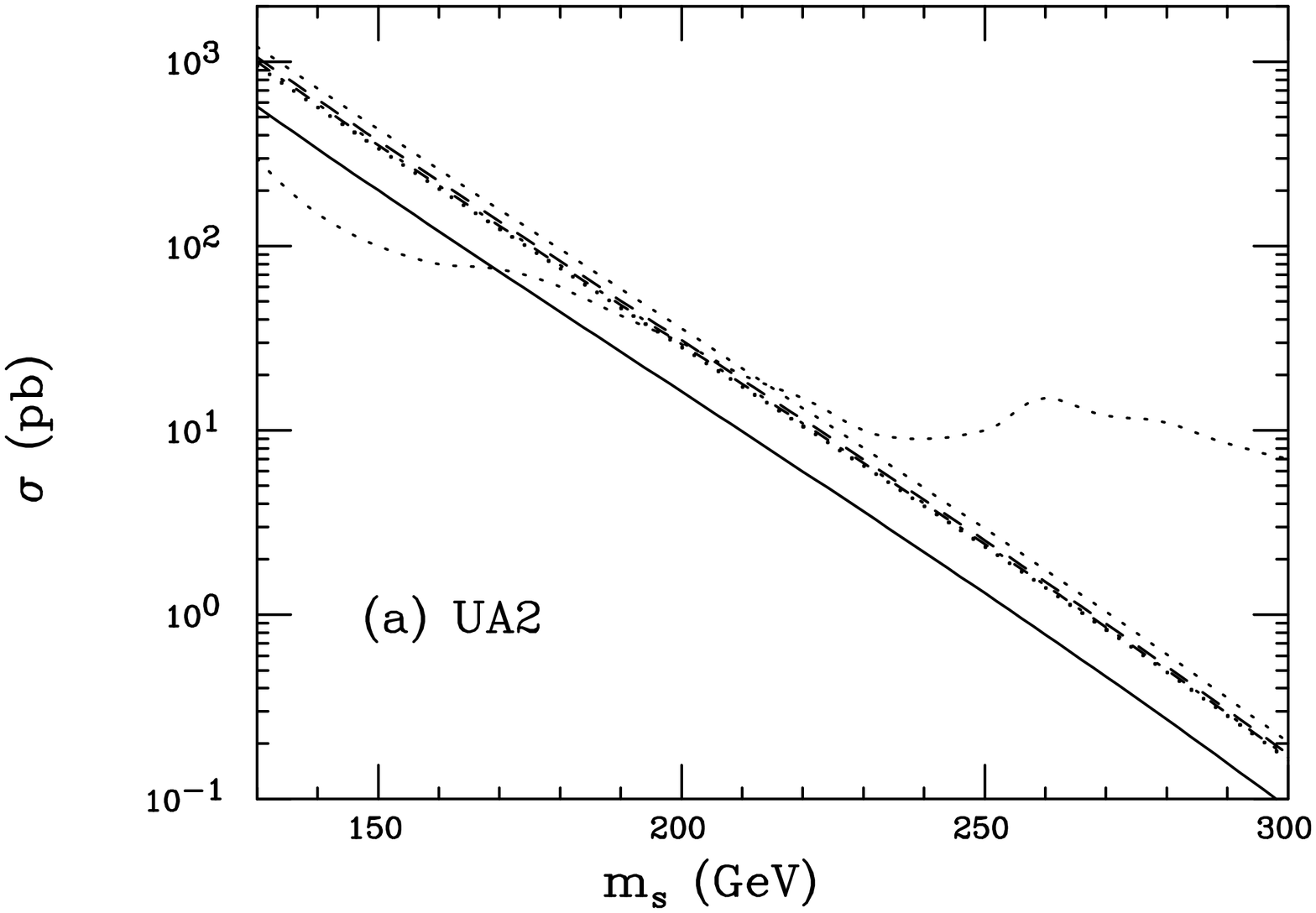,height=9cm,width=8.8cm,angle=-90}}
\vspace*{-0.75cm}
\centerline{
\psfig{figure=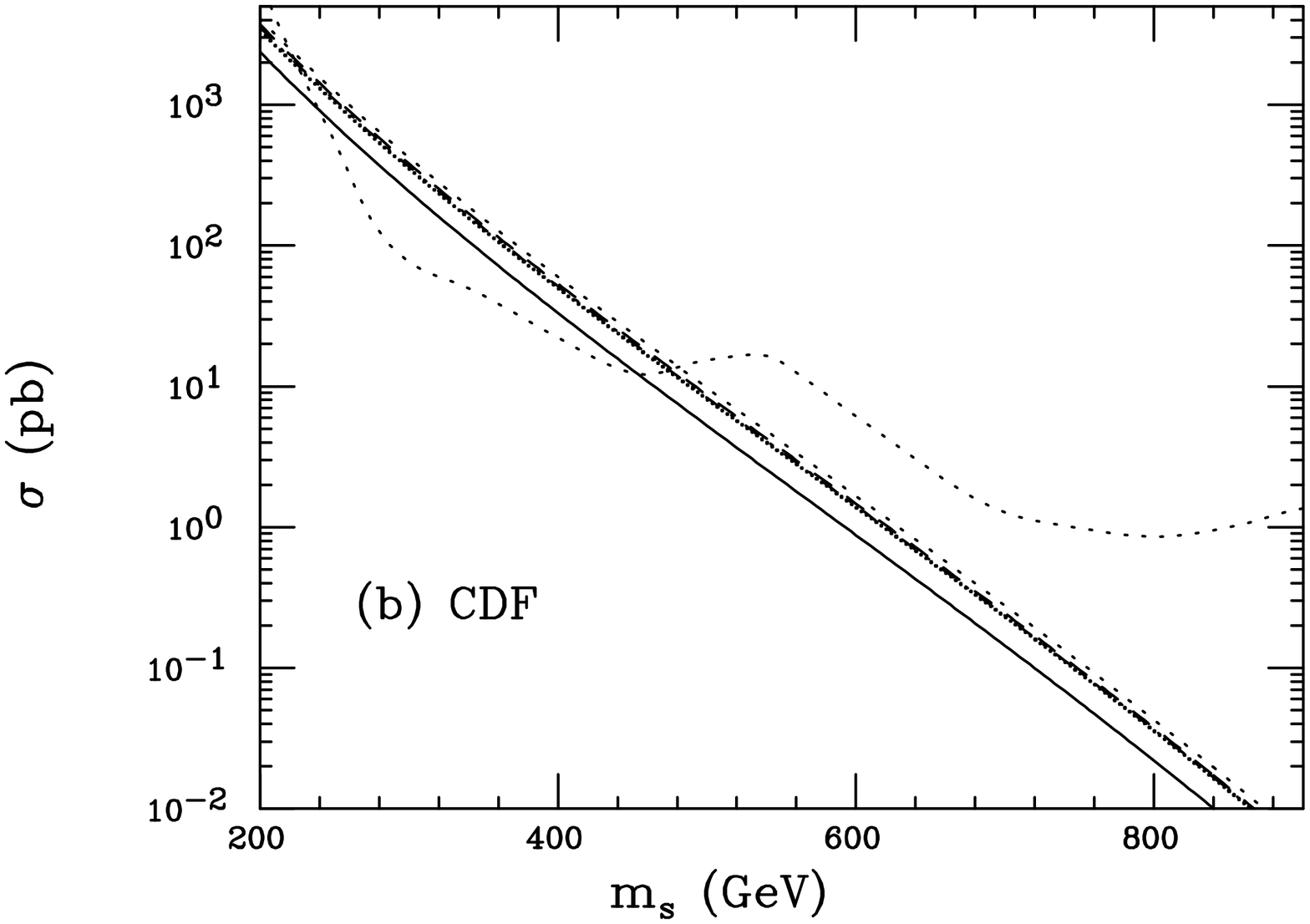,height=9cm,width=8.8cm,angle=-90}
\hspace*{-5mm}
\psfig{figure=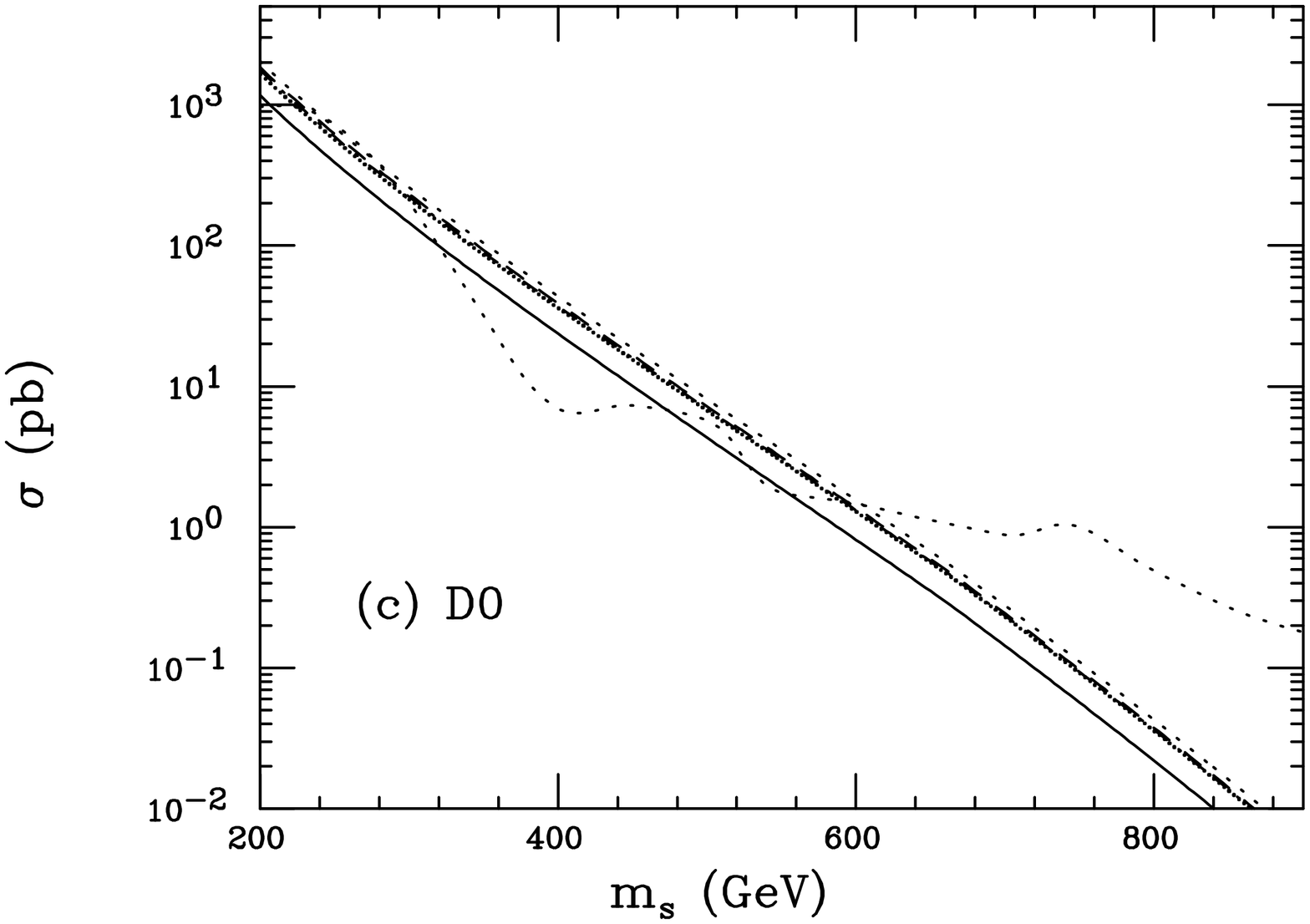,height=9cm,width=8.8cm,angle=-90}}
\vspace*{-0.75cm}
\caption{Comparison of the single squark production cross section in
the dijet channel as a function of the squark mass 
with $\mglu=0.4, 0.7, 1.1, 1.3$ and 5.0 GeV (straight curves, dotted,
dashed, dash-dotted, square-dotted, and solid from top to bottom, 
respectively)  with the upper 
bound for the production of new dijet mass resonances from 
(a) UA2 at 90\% C.L., (b) CDF and (c) D0 at 95\% C.L.  (dotted curves).}
\label{bumpsearch}
\end{figure}

Dijet angular distributions 
are a well known test of QCD and probe of new physics and have recently
been measured at the Tevatron\cite{cdfdj,d0dj}.  Ordinary QCD
processes have large $t$- and $u$-channel poles and are thus peaked in the
forward direction, whereas, resonant squark production in the light gluino
model will have a flat distribution due to the spin-0 
nature of the squark.  A convenient angular 
variable to use is $\chi\equiv \exp(|\eta_1-\eta_2|)$. 
For the case of $2\to 2$
parton scattering, this is related to the center of mass scattering angle
as $\chi=(1+|\cos\theta^*|)/(1-|\cos\theta^*|)$.  $\chi=1$ then corresponds
to $\cos\theta^*=90^\circ$.  As is well-known\cite{vb}, the advantage of 
the $\chi$ variable is that it removes the apparent singularities associated 
with the $t-$ and $u-$channel poles present in QCD.  Thus $d\sigma/d\chi$ 
shows greater sensitivity to new physics which does not possess
such poles than does $d\sigma/d\cos\theta^*$.  To show the influence of
the production of squark resonances on this distribution 
we display in Fig. 3 the ratio of $d\sigma/d\chi$ calculated  in the light 
gluino model to that of the SM, \ie, $N_\chi\equiv (d\sigma/d\chi|_{\tilde g})/
(d\sigma/d\chi|_{SM})$, for three dijet invariant mass bins
(as chosen by CDF\cite{cdfdj}) assuming a \sq\ resonance lies within each bin.
In calculating the SM distributions, we employed the MRSA$'$ parton 
densities\cite{mrsap}.  In all cases, we see that squark
production leads to an enhancement in the distribution at low values of
$\chi$ compared to the SM.  This would result in an increase in the dijet rate 
near $90^\circ$.  Comparison with
the corresponding figures presented by CDF\cite{cdfdj} shows that this
rise in $d\sigma/d\chi$ would be easily observable so that squarks
with the masses chosen here could be excluded.

\nn
\begin{figure}[htbp]
\centerline{
\psfig{figure=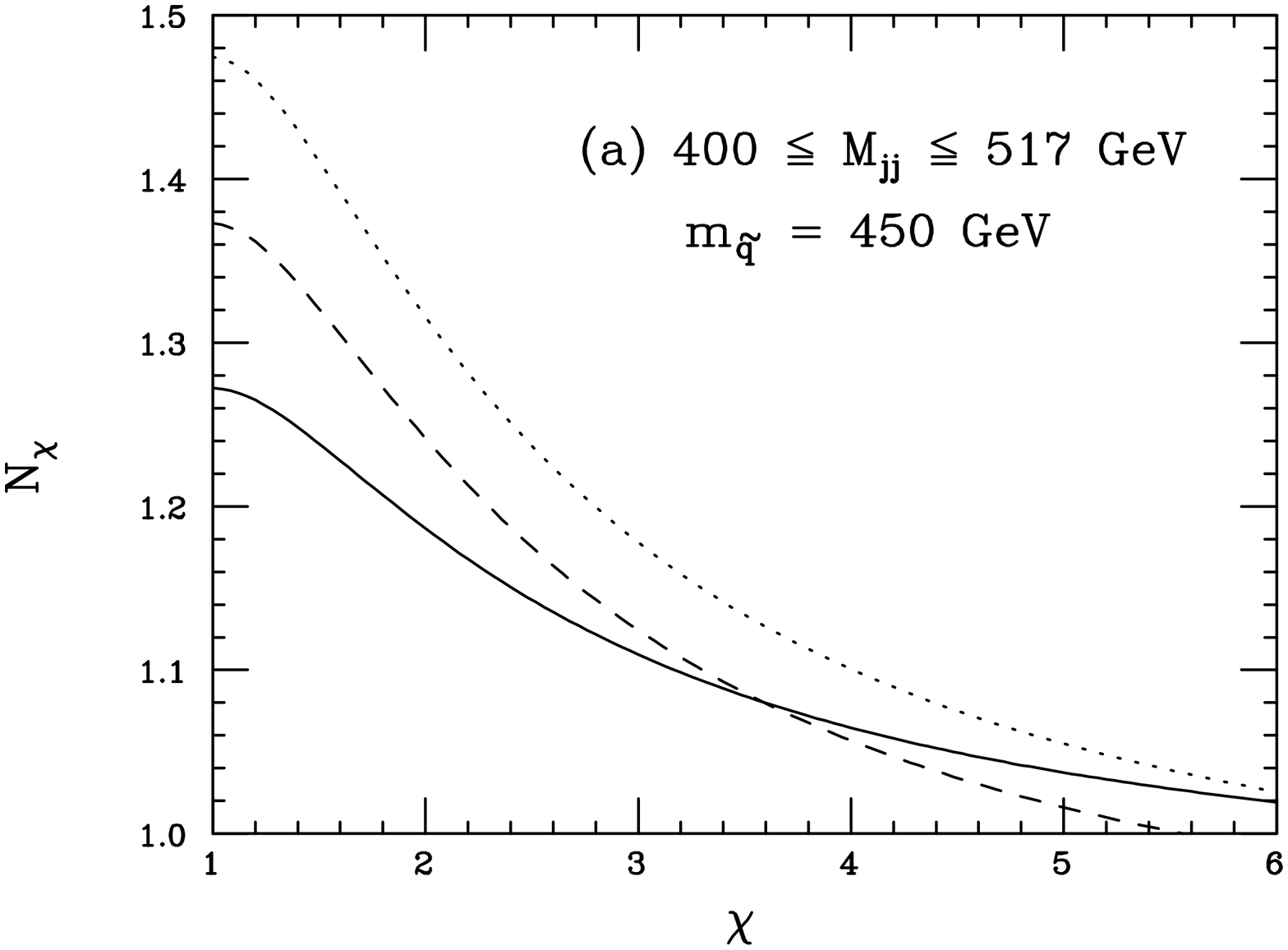,height=9.cm,width=8.8cm,angle=-90}}
\vspace*{-0.75cm}
\centerline{
\psfig{figure=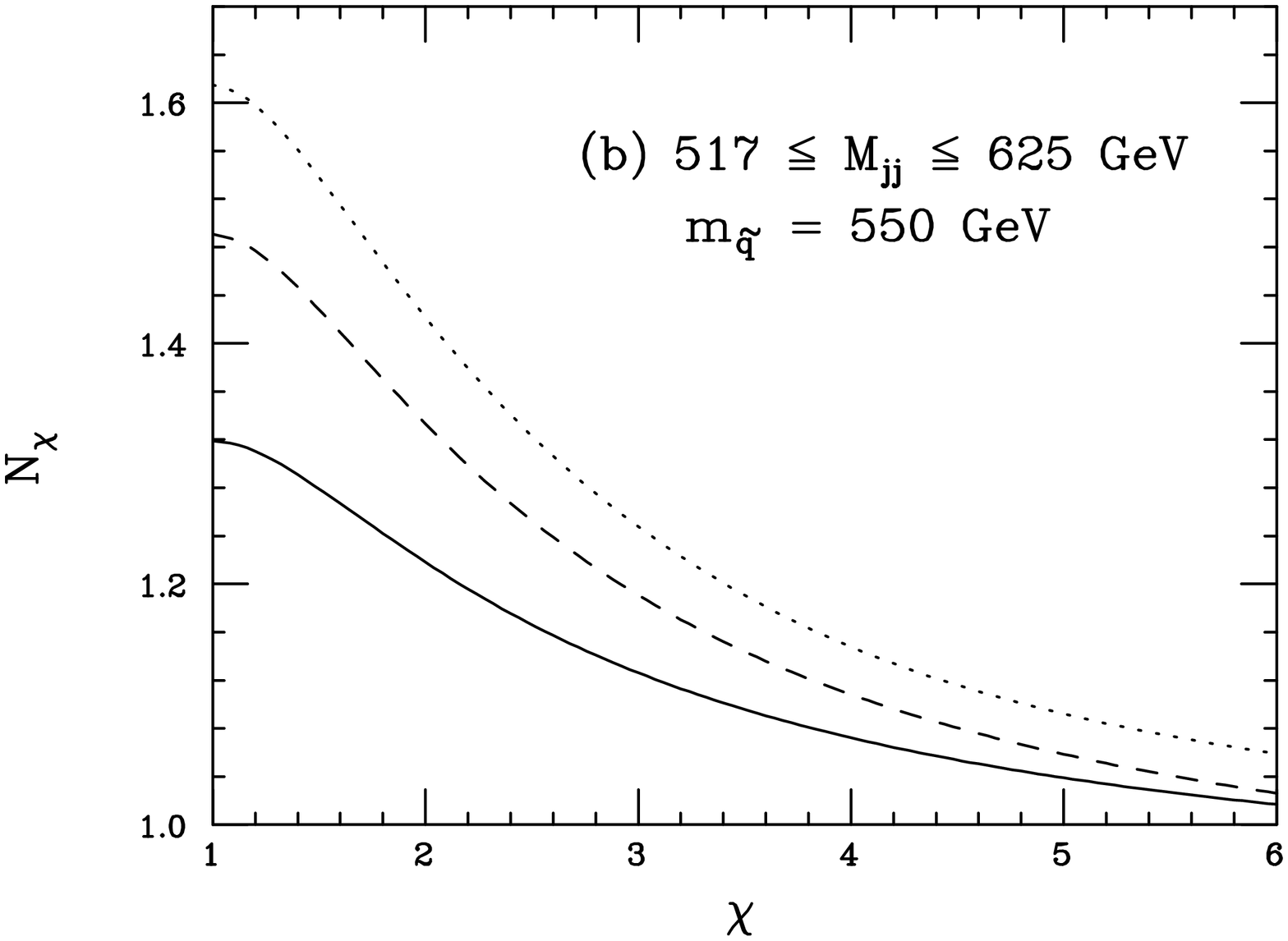,height=9.cm,width=8.8cm,angle=-90}
\hspace*{-5mm}
\psfig{figure=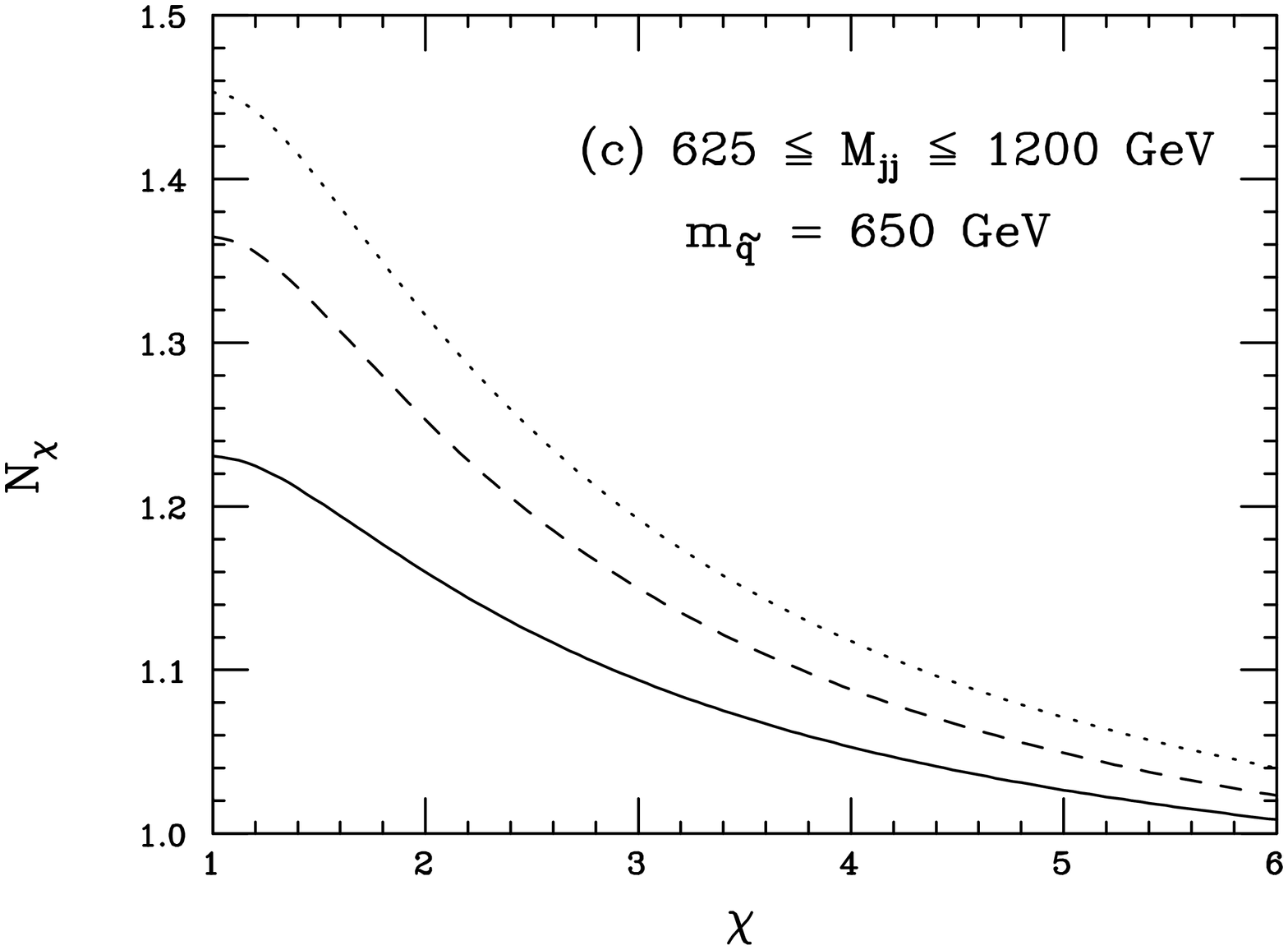,height=9.cm,width=8.8cm,angle=-90}}
\vspace*{-0.75cm}
\caption{The ratio $N_\chi\equiv (d\sigma/d\chi|_{\tilde g})/
(d\sigma/d\chi|_{SM})$ as a function of $\chi$ with gluino
masses of 0.4, 1.3, and 5 GeV, corresponding to the dotted, dashed,
and solid curves, respectively.  The dijet invariant mass bins and assumed
squark masses are as labeled.}
\label{angdist}
\end{figure}

We now make this procedure more rigorous in order to 
determine if the angular distributions can extend the
excluded regions listed in Table \ref{bumptab}.
Following the procedure used by CDF\cite{cdfdj}, we employ the variable
$R(\chi)\equiv N(\chi<2.5)/N(2.5<\chi<5)$, which is the ratio of the number
of dijet events in the two ranges of $\chi$, for the five mass bins
$241<M_{jj}<300$\,, $300<M_{jj}<400$\,, $400<M_{jj}<517$\,, $517<M_{jj}<625$, 
and $625<M_{jj}$ GeV.  This variable has the advantages that it is not
very sensitive to variations in the parton densities, to the choice of 
renormalization scale (\eg, $\mu=p_T$ versus $M_{jj}$), or to next-to-leading 
order QCD corrections, and that it characterizes the shape of the angular 
distribution in a mass bin with a single number.  We have incorporated the 
systematic errors, as determined by CDF, as well as the statistical errors
in our analysis.  The systematic errors are highly correlated, and we have
reconstructed the full covariance matrix according to the prescription in
Ref. \cite{cdfdj}.  We then calculate $R(\chi)$ in each $M_{jj}$ bin with 
$\mglu=0.4, 1.3$ and 5 GeV for squark masses in the range $160-800$ GeV, and
perform a fit to the CDF results using their data and correlation matrix.
Following the usual $\chi^2$ analysis procedure, we find the minimum value of 
$\chi^2$ for a given value of \mglu\  and then
determine the excluded range of \msq\ by examining the $\chi^2$ distribution
as a function of the squark mass. For definiteness we perform a LO calculation 
taking the scale $\mu=p_t$. Our results are presented in Fig. 4 for
each assumed value of the gluino mass. Note that the $\chi^2$ minima are 
generally found in the limit of very large squark masses. 
In all cases the $\chi^2$ distributions display a similar shape with 5 peaks 
which are associated with the 5 mass bins used by CDF and are due to the fact 
that the greatest 
sensitivity to a squark resonance occurs when it coincides in mass with the 
lower end of a given bin, \ie, when the squark cross section is maximum. 
To be more specific, when \msq\ is light ($< 241$ GeV) and outside the dijet
mass region examined by CDF, the $\chi^2$ is small but increases as the squark
mass gets closer to the edge of lowest mass bin 
and then peaks once the bin is entered. The sensitivity then decreases as \msq\
approaches the high end of the mass bin.  As the value of \msq\ rises there 
is a general loss in sensitivity 
due to decrease in statistics and the corresponding increase in the size of 
the errors. 

This analysis excludes  at the $95\%$ C.L. the \msq\ ranges
$151-694$, $166-595$, and $172-573$ GeV for \mglu=0.4, 1.3, and 5 GeV,
respectively.  It thus both extends and complements the
constraints obtained from the dijet peak searches.  Here, we might expect
improvements on these constraints for $\mglu\to 0$ due to the increased
enhancement in $N_\chi$ at $\chi=0$.  Combining these results
with the bounds from the resonance searches excludes
squark masses in the range $130<m_{\tilde q}< 694, 595, 573$ GeV 
for gluino masses of $0.4, 1.3, 5.0$ GeV.

\nn
\begin{figure}[htbp]
\centerline{
\psfig{figure=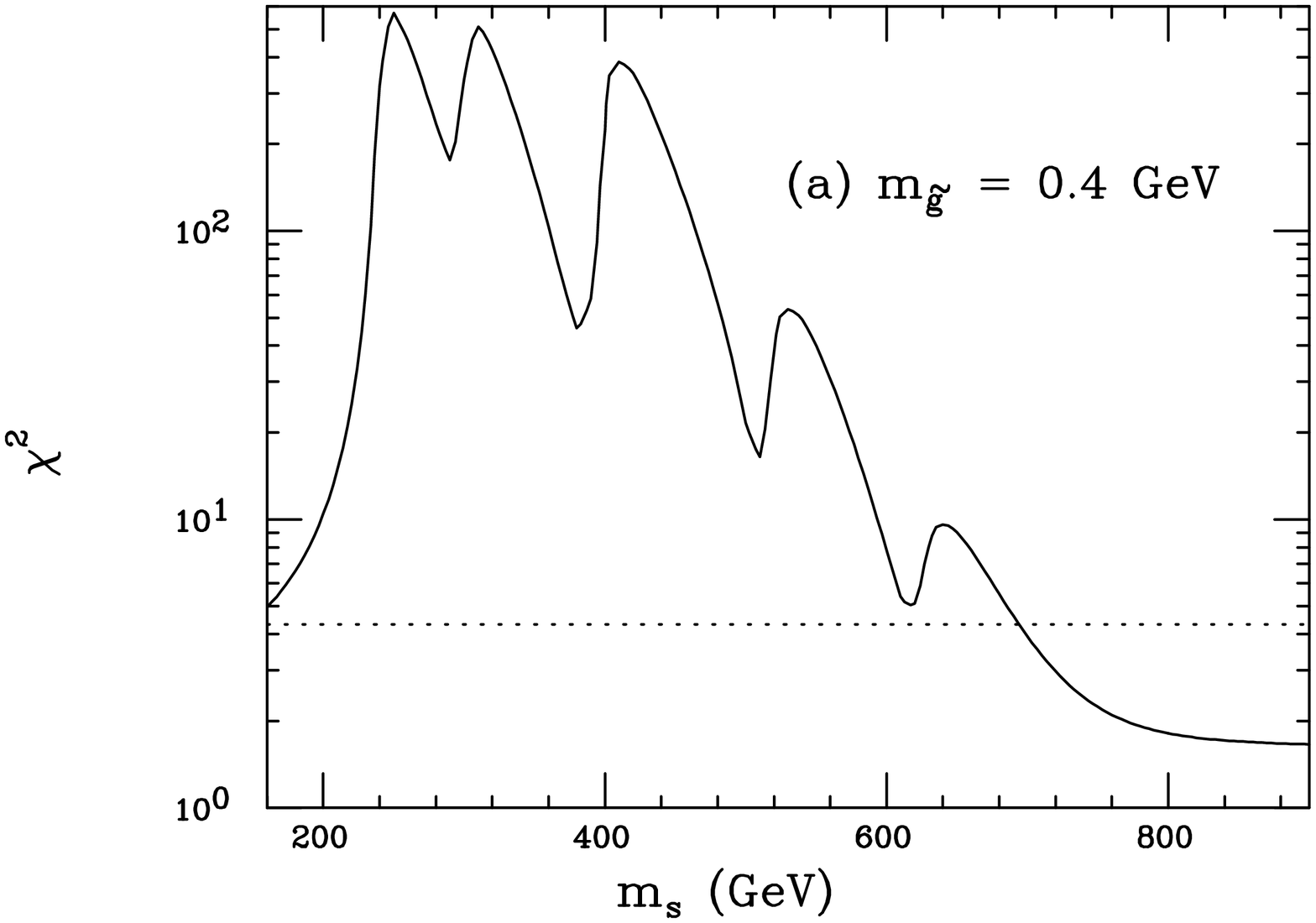,height=9.cm,width=8.8cm,angle=-90}}
\vspace*{-0.75cm}
\centerline{
\psfig{figure=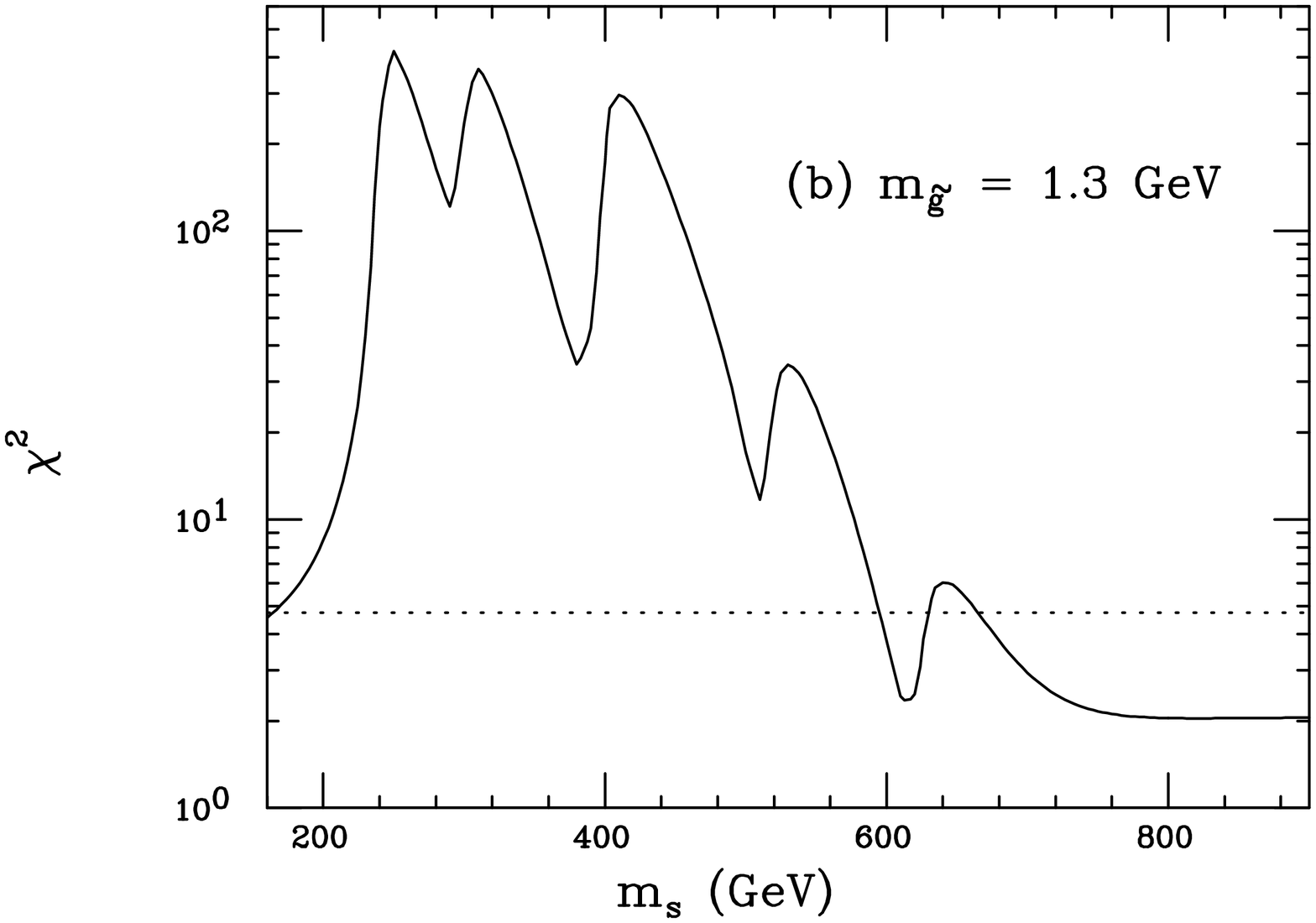,height=9.cm,width=8.8cm,angle=-90}
\hspace*{-5mm}
\psfig{figure=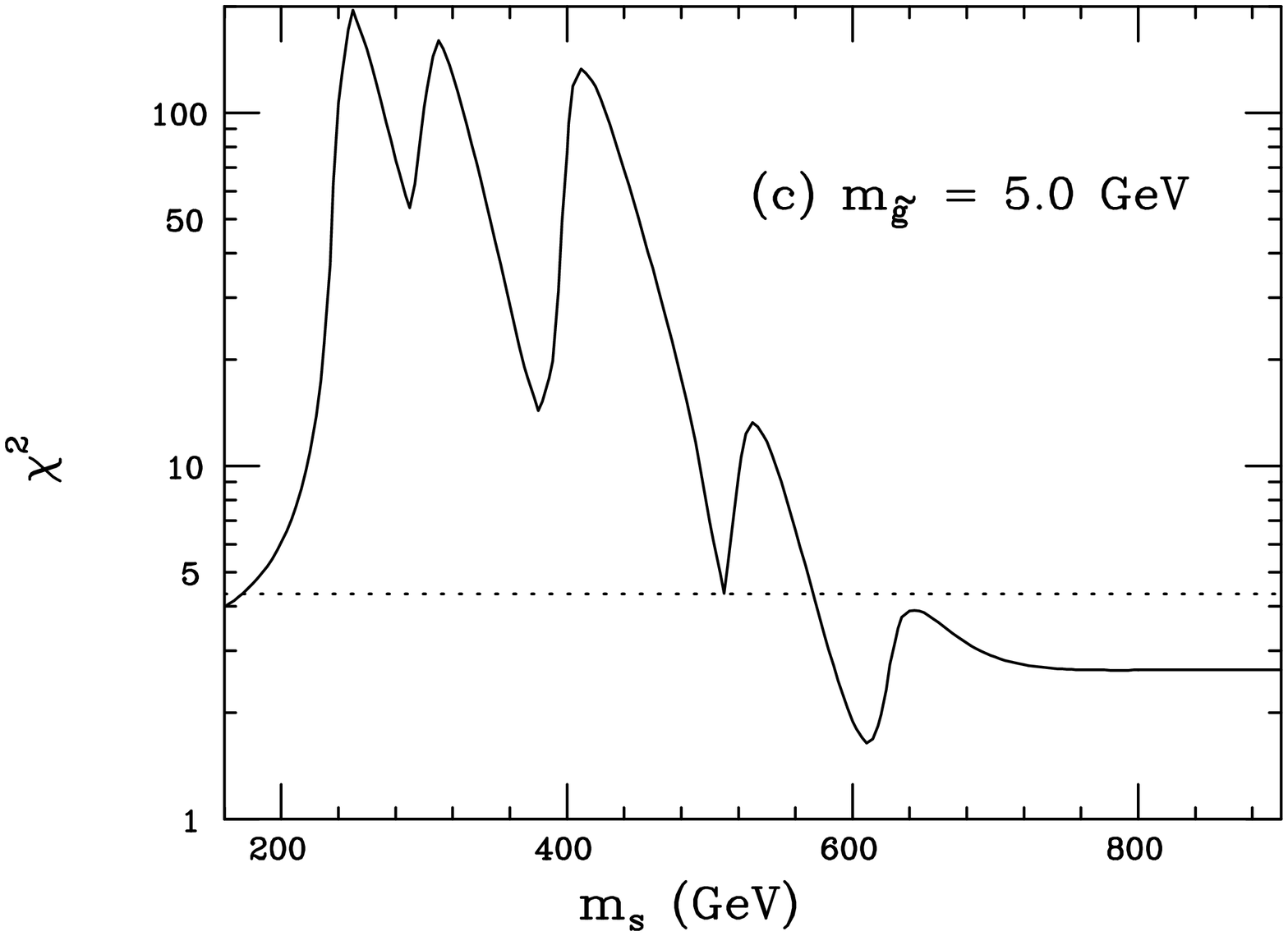,height=9.cm,width=8.8cm,angle=-90}}
\vspace*{-0.75cm}
\caption{$\chi^2$ distributions as a function of the squark mass
following the analysis described in the text, assuming gluino
masses of (a) 0.4 GeV, (b) 1.3 GeV, and (c) 5 GeV.  The dotted horizontal line
represents the $95\%$ C.L. bound in each case.}
\label{limits}
\end{figure}

In summary, we have examined the constraints on models with light gluinos
by using both the cross section and angular distribution for dijet events
observed at hadron colliders.  The critical observation is that a light
gluino can act as a partonic component of the proton thus leading to the
resonant production $q\glu\to\sq\to q\glu$, provided the \sq\ is
sufficiently light.  From our analysis, it would appear that the survival
of the light gluino case requires either a light \sq\ in the $\sim 70-130$
GeV range, or a heavy \sq\ with $\msq\gsim 600-700$ GeV.
From studies of the physics capabilities at Run II of 
the Tevatron\cite{tevstudy}, we anticipate that this future data will be 
able to exclude or verify this model for squark masses up to $\sim 1$ TeV.
High energy hadron colliders may thus provide the best testing ground for
this scenario.

\section*{Acknowledgements}

We thank both J. Stirling and A. Vogt for forwarding their numerical codes for 
the gluino parton densities, as well as R. Harris and I. Bertram for 
discussions on the CDF and D0 dijet data sample and providing us with the 
detailed numerical results from their respective experiments.

\newpage

%
\def\MPL #1 #2 #3 {Mod.~Phys.~Lett.~{\bf#1},\ #2 (#3)}
\def\NPB #1 #2 #3 {Nucl.~Phys.~{\bf#1},\ #2 (#3)}
\def\PLB #1 #2 #3 {Phys.~Lett.~{\bf#1},\ #2 (#3)}
\def\PR #1 #2 #3 {Phys.~Rep.~{\bf#1},\ #2 (#3)}
\def\PRD #1 #2 #3 {Phys.~Rev.~{\bf#1},\ #2 (#3)}
\def\PRL #1 #2 #3 {Phys.~Rev.~Lett.~{\bf#1},\ #2 (#3)}
\def\RMP #1 #2 #3 {Rev.~Mod.~Phys.~{\bf#1},\ #2 (#3)}
\def\ZPC #1 #2 #3 {Z.~Phys.~{\bf#1},\ #2 (#3)}


\begin{thebibliography}{99}

\bibitem{radmass}
R. Barbieri, L. Girardello, and A. Masiero, \PLB 127B 429 1983 ;
G.R. Farrar and A. Masiero, hep-ph/9410401.

\bibitem{pdg}
For a review see, R.M. Barnett \etal, \PRD D54 1 1996 ; G.R. Farrar,
\PRD D51 3904 1995 .

\bibitem{gfar}
G. Farrar and P. Fayet, \PLB 76B 575 1978 ; G.R. Farrar, \PRL 53 1029 1984 .

\bibitem{clavalphas}
L. Clavelli, \PRD D46 2112 1992 ; M. Jezabek and J.H. K\" uhn, 
\PLB B301 121 1993 ; J. Ellis, D.V. Nanopoulos, and D.A. Ross,
\PLB B305 375 1993 ; L. Clavelli and P.W. Coulter, \PRD D51 1117 1995 .

\bibitem{phil}
P.N. Burrows, presented at {\it 3rd International Symposium on Radiative 
Corrections}, Cracow, Poland, August 1996, hep-ex/9612007;
S. Bethke, presented at {\it QCD Euroconference 96}, Montpellier, 
France, July 1996, hep-ex/9609014.

\bibitem{cdfd0}
F. Abe \etal, (CDF Collaboration), \PRL 76 2006 1996 ; S. Abachi \etal,
(D0 Collaboration), \PRL 75 618 1995 .

\bibitem{morefar}
See, for example, G.R. Farrar, in Ref. 2, and hep-ph/9602334.

\bibitem{uaone}
C. Albajar \etal, (UA1 Collaboration), \PLB B198 261 1987 ;
R.M. Barnett, in {\it 4th International Conference on Physics Beyond the 
Standard Model}, Lake Tahoe, CA, December 1994, ed. J.F. Gunion \etal,
(World Scientific, Singapore 1995); R.M. Barnett, H.E. Haber, and G.L. Kane,
\PRL 54 1983 1985 ; V. Barger \etal, \PRD D33 57 1986 .

\bibitem{bhatt}
G. Bhattacharyya and A. Raychaudhuri, \PRD D49 R1156 1994 .

\bibitem{cusb}
P.M. Tuts \etal, (CUSB Collaboration), \PLB B186 233 1987 .

\bibitem{gfartwo}
M. Cakir and G.R. Farrar, \PRD D50 3268 1994 ; W.-Y. Keung and A. Khare,
\PRD D29 2657 1984 ; J.H. K\" uhn and S. Ono, \PLB 142B 436 1984 ;
T. Goldman and H. Haber, {\it Physica} {\bf D15}, 181 1985.

\bibitem{argus}
ARGUS Collaboration, \PLB 167B 360 1986 .

\bibitem{bdump}
J.P. Dishaw \etal, (NA3 Collaboration), \PLB 85B 142 1979 ;
F. Bergsma \etal, (CHARM Collaboration), \PLB 121B 429 1983 ; R.C. Ball \etal,
\PRL 53 1314 1984 ; WA66 Collaboration, \PLB 160B 212 1985 ; T. Akesson \etal,
(HELIOS Collaboration) \ZPC C52 219 1991 ; I.F. Albuquerque \etal,
(E761 Collaboration), hep-ex/9604002.

\bibitem{neutral}
H.R. Gustafson \etal, \PRL 37 474 1976 ; S. Dawson, E. Eichten, and C. Quigg,
\PRD D31 1581 1985 .

\bibitem{hitoshi}
R. Munoz-Tapia and W.J. Stirling, \PRD D49 3763 1994 ;
A. de Gouvea and H. Murayama, hep-ph/9606449; S. Moretti, R. Munoz-Tapia,
and J.B. Tausk, hep-ph/9609206; S. Moretti, R. Munoz-Tapia, and
K. Odagiri, hep-ph/9609235, hep-ph/9609295; DELPHI Collaboration, contributed
to the {\it 28th International Conference on High Energy Physics}, Warsaw,
Poland, July 1996, DELPHI-96-68-CONF-2; R. Akers \etal, (OPAL Collaboration),
\ZPC C65 367 1995 , and {\bf C68}, 555 (1995).

\bibitem{gfarthr}
G. Farrar, hep-ph/9608387; F. Csikor and Z. Fodor, hep-ph/9611320.

\bibitem{drees}
A. Djouadi and M. Drees, \PRD D51 4997 1995 ; B. Kileng and P. Osland,
\ZPC C66 503 1995 .

\bibitem{pdfs} 
R.G. Roberts and W.J. Stirling, \PLB 313B 453 1993 ; R. R\" uckl and A. Vogt,
\ZPC C64 431 1994 .

\bibitem{stirl}
R. Munoz-Tapia and W.J. Stirling, \PRD D52 3894 1995 .

\bibitem{zvi}
Z. Bern, A.K. Grant, and A.G. Morgan, \PLB B387 804 1996 ; L. Clavelli and
I. Terekhov, hep-ph/9605463.

\bibitem{clav}
I. Terekhov and L. Clavelli, hep-ph/9603390.

\bibitem{levan}
L.J. Clavelli and L.R. Surguladze, hep-ph/9610493.

\bibitem{mono}
D. Choudhury, hep-ph/9608444.

\bibitem{cdfdj}
F. Abe \etal, (CDF Collaboration), hep-ex/9609011.  R. Harris, in {\it
Proceedings of the 10th Topical Workshop on Proton-Antiproton Collider
Physics}, Batavia, IL 1995, ed. R. Raja and J. Yoh; we thank R. Harris for
giving us the updated version of this analysis.

\bibitem{ua2}
J. Alitti \etal, (UA2 Collaboration), \NPB B400 3 1993 .

\bibitem{d0dj}
S. Abachi \etal, (D0 Collaboration), submitted to the {it 28th International
Conference on High Energy Physics}, Warsaw, Poland, July 1996, 
FERMILAB-CONF-96-168-E. 

\bibitem{vb}
V. Barger and R.J.N. Phillips, {\it Collider Physics}, (Addison-Wesley,
Redwood City, CA, 1987).

\bibitem{mrsap}
A.D. Martin, W.J. Stirling, and R.G. Roberts, \PRD D50 6737 1994 ; 
\PLB B354 155 1995 .

\bibitem{tevstudy}
{\it Future Electroweak Physics at the Fermilab Tevatron, Report of the
Tev-2000 Study Group}, ed. D. Amidei and R. Brock, Fermilab-Pub-96/082.

\end{thebibliography}
\end{document}